\documentclass[iop]{emulateapj}
\usepackage{times,setspace,subfigure,epsf}
\usepackage{epstopdf}
\usepackage{dcolumn}
\usepackage{natbib}
\usepackage{graphicx}
\usepackage{rotating}
\usepackage{amssymb}
\usepackage{latexsym}
\usepackage{float}
\usepackage{wrapfig}
\usepackage{multirow}
\usepackage{subfigure}
\usepackage{amsmath}
\shorttitle{The Green Bank Northern Celestial Cap Pulsar Survey II: The Discovery and Timing of Ten Pulsars}

\shortauthors{Kawash et al.}

\begin{document}

\title{The Green Bank Northern Celestial Cap Pulsar Survey II: The Discovery and Timing of Ten Pulsars}

\author{A. M. Kawash\altaffilmark{1,2}, 
M. A. McLaughlin\altaffilmark{1,2}, 
D. L. Kaplan\altaffilmark{3}, 
M. E. DeCesar\altaffilmark{4},
L. Levin\altaffilmark{5},
D. R. Lorimer\altaffilmark{1,2}
R. S. Lynch\altaffilmark{6,1},
K. Stovall\altaffilmark{7},
J. K. Swiggum\altaffilmark{3}, 
E. Fonseca\altaffilmark{8}, 
A. M. Archibald\altaffilmark{9,10},
S. Banaszak\altaffilmark{3},
C. M. Biwer\altaffilmark{11},
J. Boyles\altaffilmark{12},
B. Cui\altaffilmark{1,2}, 
L. P. Dartez\altaffilmark{13},
D. Day\altaffilmark{13},
S. Ernst\altaffilmark{14},
A. J. Ford\altaffilmark{13},
J. Flanigan\altaffilmark{3},
S. A. Heatherly\altaffilmark{3}
J. W. T. Hessels\altaffilmark{9,10},
J. Hinojosa\altaffilmark{13},
F. A. Jenet\altaffilmark{13},
C. Karako-Argaman\altaffilmark{8},
V. M. Kaspi\altaffilmark{8},
V. I. Kondratiev\altaffilmark{9,15},
S. Leake\altaffilmark{13},
G. Lunsford\altaffilmark{13},
J. G. Martinez\altaffilmark{16},
A. Mata\altaffilmark{13},
T. D. Matheny\altaffilmark{1,2},
A. E.  Mcewen\altaffilmark{1,2},
M. G. Mingyar\altaffilmark{1,2},
A. L. Orsini\altaffilmark{1,2},
S. M. Ransom\altaffilmark{17},
M. S. E. Roberts\altaffilmark{18},
M. D. Rohr\altaffilmark{3},
X. Siemens\altaffilmark{3},
R. Spiewak\altaffilmark{19,3},
I. H. Stairs\altaffilmark{20},
J. van Leeuwen\altaffilmark{9,10},
A. N. Walker\altaffilmark{3},
B. L. Wells\altaffilmark{21,3},
}

\altaffiltext{1}{Center for Gravitational Waves and Cosmology, Chestnut Ridge Research Building, Morgantown, WV 26505}
\altaffiltext{2}{Department of Physics and Astronomy, West Virginia University, Morgantown, WV 26506}
\altaffiltext{3}{Center for Gravitation, Cosmology, and Astrophysics, Department of Physics, University of Wisconsin-Milwaukee, PO Box 413,
Milwaukee, WI, 53201-0413, USA}
\altaffiltext{4}{Department of Physics, 730 High St., Lafayette College, Easton, PA, 18042-1761, USA}
\altaffiltext{5}{Jodrell Bank Centre for Astrophysics, School of Physics and Astronomy, The University of Manchester, Manchester, M13 9PL, UK}
\altaffiltext{6}{Green Bank Observatory, PO Box 2, Green Bank, WV 24494-0002, USA}
\altaffiltext{7}{National Radio Astronomy Observatory, 1003 Lopezville Rd., Socorro, NM, 87801-4867}
\altaffiltext{8}{Department of Physics and McGill Space Institute, McGill University, 3600 University Street, Montreal, QC, H3A 2T8, Canada}
\altaffiltext{9}{ASTRON, the Netherlands Institute for Radio Astronomy, Postbus 2, 7990 AA, Dwingeloo, The Netherlands}
\altaffiltext{10}{Anton Pannekoek Institute for Astronomy, University of Amsterdam, Science Park 904, 1098 XH, Amsterdam, The Netherlands}
\altaffiltext{11}{Department of Physics, Syracuse University, Syracuse, NY, 13244-0001, USA}
\altaffiltext{12}{Department of Physics and Astronomy, Western Kentucky University, 1906 College Heights Blvd., Bowling Green, KY, 42101-1000, USA}
\altaffiltext{13}{Center for Advanced Radio Astronomy, University of Texas Rio Grande Valley, 1 W. University Blvd., Brownsville, TX, 78520-4933, USA}
\altaffiltext{14}{Space Science Center, Department of Earth and Space Sciences, Morehead State University, 235 Martindale Drive, Morehead, KY 40351}
\altaffiltext{15}{Astro Space Center of the Lebedev Physical Institute, Profsoyuznaya str. 84/32, Moscow 117997, Russia}
\altaffiltext{16}{Max-Planck-Institut f\"{u} Radioastronomie, Auf dem H\"{u}gel 69, D-53121, Bonn, Germany}
\altaffiltext{17}{National Radio Astronomy Observatory, 520 Edgemont Road, Charlottesville, VA, 23903, USA}
\altaffiltext{18}{Eureka Scientific, Inc., 2452 Delmer St., Suite 100, Oakland, CA, 94602-3017, USA}
\altaffiltext{19}{Centre for Astrophysics and Supercomputing, Swinburne University of Technology, PO Box 218, Hawthorn, VIC 3122, Australia}
\altaffiltext{20}{Department of Physics and Astronomy, University of British Columbia, 6224 Agriculture Rd., Vancouver, BC, V6T 1Z1, Canada}
\altaffiltext{21}{Department of Atmospheric Sciences, Colorado State University, 3915 W. Laporte Ave., Fort Collins, CO, 80523-0001}

\begin{abstract}

We present timing solutions for ten pulsars discovered in 350 MHz searches with the Green Bank Telescope. Nine of these were discovered in the Green Bank Northern Celestial Cap survey and one was discovered by students in the Pulsar Search Collaboratory program in analysis of drift-scan data. Following discovery and confirmation with the Green Bank Telescope, timing has yielded phase-connected solutions with high precision measurements of rotational and astrometric parameters. Eight of the pulsars are slow and isolated, including PSR J0930$-$2301, a pulsar with nulling fraction lower limit of $\sim$30\% and nulling timescale of seconds to minutes. This pulsar also shows evidence of mode changing. The remaining two pulsars have undergone recycling, accreting material from binary companions, resulting in higher spin frequencies. PSR J0557$-$2948 is an isolated, 44 \rm{ms}  pulsar that has been partially recycled and is likely a former member of a binary system which was disrupted by a second supernova.  The paucity of such so-called `disrupted binary pulsars' (DRPs) compared to double neutron star (DNS) binaries can be used to test current evolutionary scenarios,  especially the kicks imparted on the neutron stars in the second supernova. There is some evidence that DRPs have larger space velocities, which could explain their small numbers. PSR J1806+2819 is a 15 \rm{ms} pulsar in a 44 day orbit with a low mass white dwarf companion. We did not detect the companion in archival optical data, indicating that it must be older than 1200 Myr.  

\end{abstract}

\section{Introduction}
\label{intro}

The Green Bank Northern Celestial Cap (GBNCC) survey is a 350 MHz all-sky pulsar survey being conducted with the Green Bank Telescope (GBT; \citealt{slr14}). The primary goals of the survey include the discovery of millisecond pulsars (MSPs) suitable for pulsar timing arrays (PTAs) for direct detection of gravitational wave (GW; \citealt{abb18,ltm15,srm13}), exotic pulsar systems such as `black widows' \citep{fst88} and `redbacks' \citep{r11,asr09}, rotating radio transients (RRATs; \citealp{mll06,dsm16}), and characterization of the Galactic pulsar population. 

This survey is 75\% complete and will ultimately cover the entire sky visible to the GBT ($\delta>-40^\circ$)  with $\sim$125000 pointings, 120 \rm{s} each. To date, there have been 156 pulsars, including 20 MSPs, discovered in this survey\footnote[1]{http://astro.phys.wvu.edu/GBNCC/}. Collection of data is expected to be complete in 2020. Lynch et al. 2018 (submitted) report on the discovery of 45 pulsars, including five millisecond pulsars (MSPs), a new relativistic double neutron star system, an intermediate mass binary pulsar, a mode-changing pulsar, a non-recycled pulsar with a very low magnetic field, and several nulling pulsars. Here, we report on the discovery of nine pulsars in the GBNCC survey, including a millisecond pulsar, a partially recycled pulsar, and a nulling pulsar.

The Pulsar Search Collaboratory (PSC)\footnote[2]{http://pulsarsearchcollaboratory.com/} is a joint outreach program between West Virginia University (WVU) and the Green Bank Observatory (GBO). The primary goals of the PSC are to stimulate high-school students' interest in science and to discover unknown pulsars. Since its start in 2008, the PSC has integrated high-school students into the pulsar searching process resulting in the discovery of seven pulsars \citep{rsm13,srm15}. PSR J1954$+$1021 is the seventh pulsar discovered by the PSC, and its timing solution is presented here, in addition to those of the nine newly discovered GBNCC survey pulsars.

To date, 9\% of the $\sim$2600 known pulsars are in binary systems \citep{mhth05}. These include pulsars orbiting white dwarf (WD), neutron star (NS), and main sequence (MS) companions. Pulsars with WD and NS companions are likely recycled through the following process. From a binary star system, the more massive star will undergo a supernova explosion and form a pulsar. Estimating using the virial theorem, the supernova explosion will cause the binary star system to be disrupted if more than half the pre-supernova mass is ejected from the system \citep{h83,bh91}. The presence of a kick could change this requirement significantly. If the system remains bound, and the companion is massive enough to evolve into a giant and overflow its Roche lobe, material is accreted onto the pulsar. This process, commonly referred to as `recycling,' transfers angular momentum to the pulsar and spins it up \citep{acrs82}.

The amount and timescale of recycling depends on the binary companion type. MSPs, which undergo the greatest amount of recycling, usually have WD companions \citep{tlm12}. However, a surprising number are isolated, presumably due to ablation of their companion stars \citep{fst88} or three-body interactions \citep{phl11}. Partially-recycled pulsars, with spin periods $P$ and period derivatives $\dot{P}$ in the 30 \rm{ms} \textless $P$ \textless  100 \rm{ms} and $\dot{P}$ \textless  10$^{-17}$ range, typically have more massive WD or NS companions \citep{l08}. Some are also isolated due to the binary system being disrupted during the supernova explosion of the secondary \citep{cnt93}.  

Two of the pulsars presented in this paper have undergone a phase of recycling, presumably in a binary evolution scenario described above. PSR J0557$-$2948 is a partially-recycled pulsar that has no evidence of a companion star and is discussed in Section \ref{0557}. PSR J1806+2819 is a recycled pulsar in a binary system that is discussed in further detail in Section \ref{1806}.

\section{Methodology and Results}
\label{results}

A description of the survey and the searching pipeline are given in \cite{slr14}. The search processing responsible for these discoveries took place on the Guillimin cluster operated by CLUMEQ and Compute Canada and the candidate plots were analyzed using the CyberSKA interface\footnote[3]{ https://ca.cyberska.org/}. Timing observations of the pulsars presented in this paper were conducted with the GBT using a combination of competitively awarded GBNCC timing time and time purchased by WVU. The observations were typically 15 to 20 minutes in duration.

Using the Green Bank Ultimate Pulsar Processing Instrument (GUPPI; \citealt{rdf09}), data were recorded over 100 MHz of 
bandwidth divided in to 4096 frequency channels at a central frequency of 350 MHz every 81.92 
$\mu$s.  In off-line processing, the filterbank data were dedispersed at the dispersion measure (DM) of the pulsar to account for frequency-dependent time delays due to the interaction of the pulse and the 
interstellar medium. 

\begin{figure*}
         \begin{center}
                \includegraphics[width=1.0\textwidth]{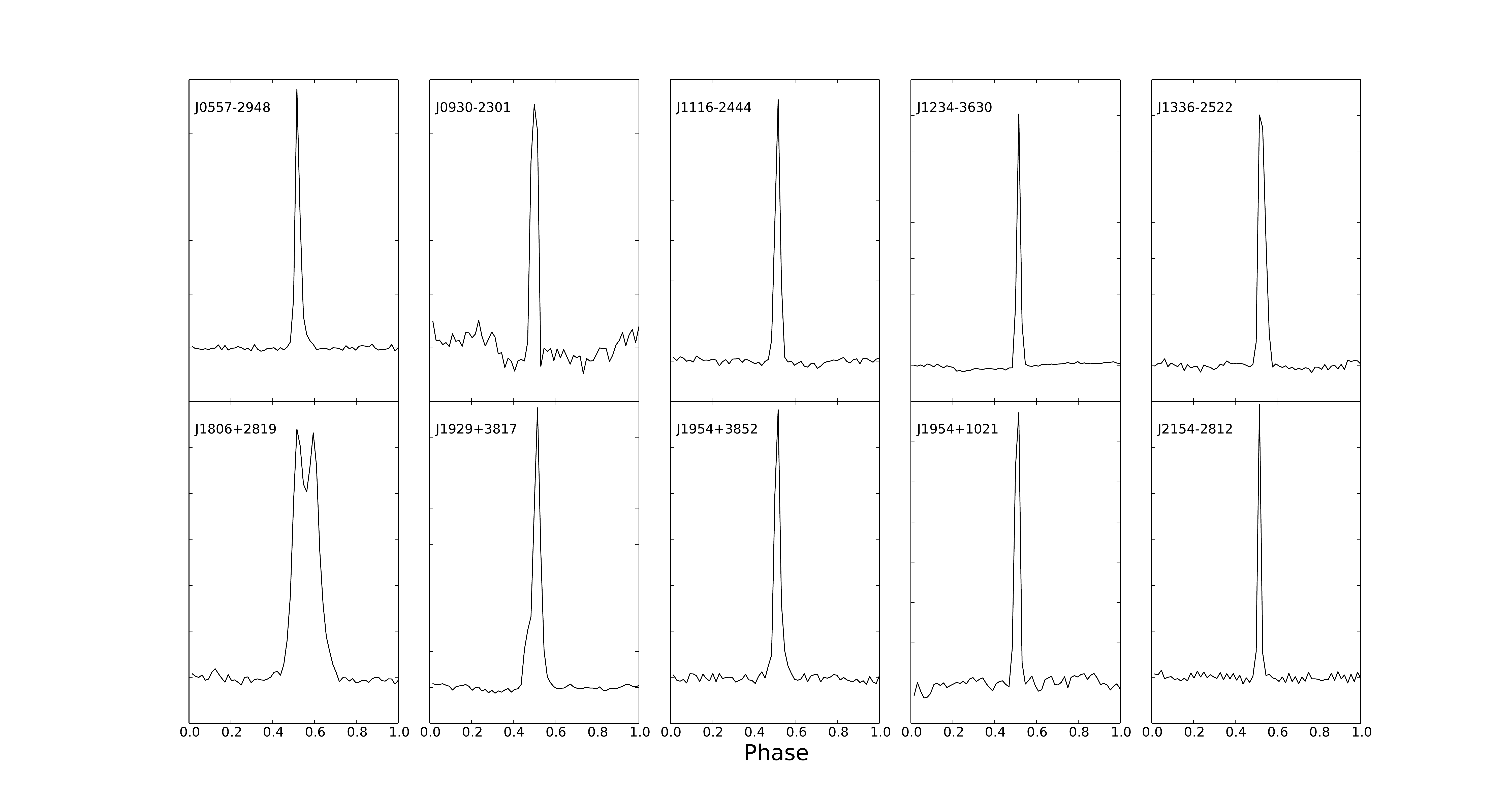}
           \caption{Integrated pulse profiles at a central frequency of 350 MHz. The horizontal axis shows pulse phase and the vertical axis shows intensity. These profiles were formed by summing pulses modulo the pulse period using 64 phase bins at each epoch and averaging the profiles from all of the epochs weighted by the signal-to-noise ratio ($S/N$).}
        \label{fig:profiles}
        \end{center}
\end{figure*}

To calculate the pulse times of arrival (TOAs) at the telescope, integrated profiles were formed by summing many 
thousands of pulses modulo the pulse period at each epoch. The pulse profile for each of the pulsars can be found in Figure \ref{fig:profiles}. These pulse profiles were cross-correlated by the least squares method to template profiles, created by summing together profiles at multiple epochs, to achieve a high precision TOA measurement of the pulse at the telescope \citep{t93}.  TOAs were calculated using \texttt{get{\_}TOAs.py} in the PRESTO software package\footnote[4]{http://www.cv.nrao.edu/$\sim$sransom/presto/} \citep{rs11}.

A timing model incorporating $P$,  $\dot{P}$, right ascension (RA), and declination (Dec.) was then fit to the TOAs using the TEMPO software package\footnote[5]{http://tempo.sourceforge.net}. This yields a timing solution that accounts for every rotation of the pulsar over the entire data span. The results of these fits are shown in Tables 1 and 2. Each of our timing solutions uses the DE430 Solar System ephemeris \citep{fwb14} and the Terrestrial Time scale TT(BIPM). The uncertainties in TOAs were scaled using EFACs to assure that the chi-squared value for each fit was equal to one. A full phase-connected timing solution reveals how well the timing model can predict when the next pulse will arrive at the telescope. The timing residuals are the measured difference between the observed and predicted arrival times. The RMS values of the timing residuals are listed in Table \ref{tb:prd}.

The DMs listed in Table \ref{tb:posn} were calculated for eight of the pulsars by creating TOAs in four subbands of the 100 MHz bandpass at one epoch. The frequency-dependent time delay of the pulse was then modeled as pulse dispersion using the TEMPO software package. This yielded a measurement of the DM of the pulsar at that epoch. PSRs J0930$-$2301 and J1954$+$1021 were observed at a central frequency of 820 MHz and bandwidth of 200 MHz in addition to the 350 MHz central frequency observations. This allowed the DM to be measured between epochs observed at different frequencies. Figure 2 shows these pulsars on the $P$--$\dot{P}$ diagram with all of the known Galactic pulsars (i.e. excluding pulsars found in globular clusters). As seen in this Figure, the timing-derived parameters for the pulsars presented in this paper lie within the range of parameters for known Galactic pulsars.

\begin{table*}
\label{tb:posn}
\begin{center}
 \caption{Timing Derived Positions and Distances}
\begin{footnotesize} \begin{tabular}{lllrrlll}
\hline Name & R.A. (J2000) & Dec. (J2000) & {$l$} & {$b$} & DM & Distance$^a$ & Distance$^b$\\ & (h~~~m~~~s) & (\degr ~~~\arcmin ~~~\arcsec) & {(\degr)} & {(\degr)} & (pc cm$^{-3}$) & (kpc) & (kpc) \\
\hline
J0557$-$2948  & 05:57:32.9995(7) & $-$29:48:16.804(7) & 235.5 & $-$23.99 & 49.05(1) & 4.3 & 2.9\\
J0930$-$2301  & 09:30:02.82(8) & $-$23:01:45(1) & 253.9  & 20.14  & 78.3(5) & $>$25$^c$ & $>$50$^c$\\ 
J1116$-$2444  & 11:16:23.26(3) & $-$24:44:56.1(5) & 277.0 & 33.29 & 29.8(3) & 0.9 & 1.2\\
J1234$-$3630  & 12 34 12.00(2) & $-$36 30 41.1(1) & 299.1 & 26.23  & 58.8(4) & 5.7  & 2.5\\
J1336$-$2522  & 13:36:20.84(2) & $-$25:22:01.6(5) &  315.6 & 36.40 & 37.5(2) & 5.5 & 1.7 \\
J1806$+$2819  & 18:06:25.0658(5) & \phantom{$+$}28:19:01.115(5) &  54.6 & 21.67 &  18.6802(4) & 1.3 & 1.3\\
J1929$+$3817  & 19:29:07.014(5) & \phantom{$+$}38:17:57.5(1) & 71.2 & 9.69 & 93.4(2) & 9.3 & 5.1 \\
J1954$+$3852  & 19:54:01.083(4) & \phantom{$+$}38:52:15.88(5) & 74.0 & 5.70 & 65.4(1) & 4.7 & 3.7\\
J1954$+$1021  & 19:54:36.80(1) & \phantom{$+$}10:21:10.5(9) &  49.5 & $-$8.99 & 80.87(4) & 4.3 & 3.6  \\
J2154$-$2812  & 21:54:17.37(7) & $-$28:12:41(1) & 82.0 & $-$20.16 & 32.1(9) & 2.7 &  2.0 \\ \hline \end{tabular}
\end{footnotesize}\end{center}
{\small {\bf Notes.} Timing-derived positions, Galactic longitudes and latitudes, DMs, and distances. The numbers in parentheses after position and DM are the 1-$\sigma$ errors in the last digit reported by TEMPO. 

$^a$ Distances derived from the YMW16 \citep{ymw17} electron density model.

$^b$ Distances derived from the NE2001 \citep{cl02} electron density model.

$^c$ The DM value for this pulsar is larger than the expected highest DM given by both the YMW16 and the NE2001 models, and, therefore, these models do not provide a reliable distance estimate for this source.}\end{table*}

\begin{table*}
\label{tb:prd}
\begin{center}
\caption{Timing Solutions and Derived Parameters}
\begin{footnotesize} \begin{tabular}{lllllllllllll}
\hline 
Name & $P$ & $\dot{P}$ & Epoch & $w_{50}$ & RMS &  Data Span & Age & $B$ &  $\dot{E}$  \\
 & (s)   & (10$^{-15}$) & (MJD) & (ms) & ($\mu$s) & (MJD) & (Myr) &  (10$^{9}$G)  & (10$^{30}$ erg s$^{-1}$) \\
\hline
J0557$-$2948 & 0.0436426389000(3) & 0.000073(5) & 57381 & 1.1 & 23.3 &  57062--57700 & 9400  & 1.8 & 35  \\
J0930$-$2301 & 1.80706867799(8) & 3.362(7) & 57254 &  62.1 & 3586.1 & 56663--57846 & 8.5 & 2500 & 22  \\ 
J1116$-$2444 & 0.86794888009(5) & 0.985(5) & 57513 & 24.3 & 736.0  &  57213--57813 & 14 & 950 & 61\\
J1234$-$3630 & 0.569242225079(8) & 0.866(3) & 57602 & 11.2 & 259.1 & 57385--57819 & 10. & 710 & 190  \\
J1336$-$2522 & 0.478145482800(6) & 0.3306(5) & 57514 & 19.0 & 237.0  & 57215--57813 & 24 & 400 & 120 \\
J1806$+$2819 & 0.01508366732422(2) &0.0000375(5) & 57004 & 2.0 & 16.2 & 56254--57753 & 6400 & 0.76 & 430  \\
J1929$+$3817 & 0.81421524225(1)& 0.6097(9) & 57545 & 29.7 &  206.7 & 57216--57875  & 22 & 700 & 43 \\
J1954$+$3852 & 0.352933478726(1) & 6.5998(1) & 57485 & 11.0 & 166.5 &  57094--57875 & 0.85 & 1500 & 5900  \\
J1954$+$1021 & 2.09944017034(5) & 1.735(3) & 57393 & 65.2 & 934.1 & 56911--57875 & 19 & 1900 & 7.4  \\ 
J2152$-$2812 &  1.3433614881(1) & 0.613(8) & 57297 & 23.1 & 1615.5 & 56838--57756 & 35 & 920 & 1.0  \\ \hline \end{tabular}
\end{footnotesize}\end{center}
{\small {\bf Notes.} Periods, period derivatives, MJDs of the epoch used for the period determination, the average pulse widths at 50\% of the peak, the RMS values of the post-fit timing residual, the MJD ranges covered, the spin-down age, surface magnetic field strength, and spin-down luminosity. No attempt was made to correct for bias in the observed spin period derivative, $\dot{P}$, by accounting for Galactic acceleration \citep{nt95} or proper motion \citep{s70,dt92}. The numbers in parentheses after $P$ and $\dot{P}$ are the 1-$\sigma$ errors in the last digit reported by TEMPO. All timing solutions use the DE430 Solar System Ephemeris.}
\end{table*}

\section{Discussion}
\label{conclusions}
Here, we discuss in further detail the most interesting pulsars presented in this paper. In Section \ref{0557} we discuss PSR J0557$-$2948 and the discrepancy in the number of predicted and known disrupted recycled pulsars to double neutron star systems. Section \ref{1806} discusses the measurement of the binary parameters of PSR J1806+2819, the search for an optical counterpart, and the possibility of including this MSP in PTAs. In Section \ref{0930} we explain the nulling analysis performed on PSR J0930$-$2301. We also discuss the significant differences between the distances derived by two different electron density models for all ten pulsars in Section \ref{models}.

\begin{figure}
         \begin{center}
                \includegraphics[width=0.5\textwidth,angle=270]{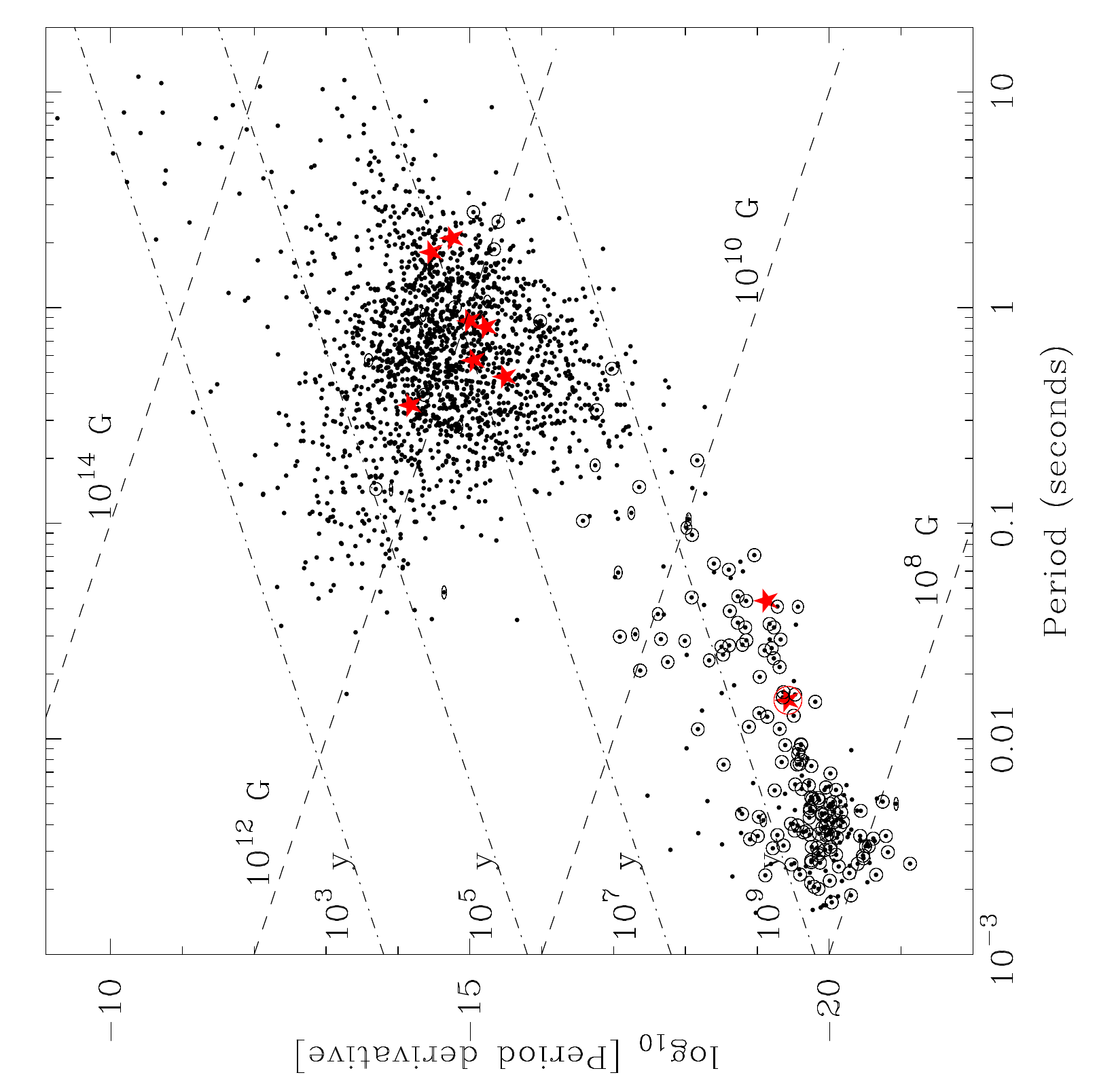}
        \label{fig:ppdot}
        \end{center}
\caption{The $P$--$\dot{P}$ diagram for known pulsars marked by dots and the pulsars with new timing solutions marked by red stars. Constant characteristic age and constant inferred surface dipole magnetic field strength are indicated by dot-dashed lines and dashed lines, respectively. Ellipses around dots and stars indicate a binary pulsar, with the ellipticity equal to the orbital ellipticity.} 
\end{figure}

\subsection{PSR J0557$-$2948}
\label{0557}
PSR J0557$-$2944 is a partially recycled pulsar with a spin period of 43.6 \rm{ms} and a DM of 49 pc cm$^{-3}$. The timing solution indicates that this pulsar has undergone a phase of recycling  but is now isolated. More specifically, this pulsar has a low period derivative ($\dot{P} = 7.3 \times10^{-20}$) and the spin period shows no evidence of a periodic Doppler shift. This pulsar is likely the end result of a disrupted neutron star (DNS) system. This class of neutron stars began to be identified in the 1990s \citep{cnt93} and have since been dubbed disrupted recycled pulsars \citep[DRPs;][]{lma04,h83,bh91,tt98}.

An interesting diagnostic of binary neutron star evolution scenarios is the relative numbers and distribution
of DRPs and DNSs in the Galaxy. As discussed by previous authors \citep[see, e.g.,][]{lma04}, the number of
DRPs relative to DNSs will depend on the survival probability of the second supernova explosion that formed
these systems: higher survival probabilities (for example, due to relatively runaway velocities) will increase
the numbers of DNS binaries relative to DRPs. It is also possible that DRPs actually have larger runaway velocities, and, hence, are fainter, and/or they escape the galaxy more frequently than bound systems. 
 
To investigate the latter hypothesis, we examined the height ($z$)
from the Galactic plane,
flux and
luminosity distributions of DRPs and DNSs.
Given the small numbers of both types of pulsars, statistical estimates
such as the Kolmogorov-Smirnov test have limited ability to determine
whether there are any significant differences between the two populations.
From a simple comparison based largely on DM-derived distances, we note that median and mean $z$-heights
are 200~pc and 300$\pm$100~pc for DNS binaries versus 385~pc and 580$\pm$160~pc
for DRPs. While this appears to support the idea that DRPs have higher
space velocities, we find no significant difference in the luminosity or
flux distributions for the two classes. The lack of any difference in
the fluxes is surprising, since for a common luminosity distribution to both
sources, one would naively expect a lower average flux from the DRPs if their
scale heights were truly larger. A more detailed observational and simulation
study, which considers the selection biases in both DRPs and DNS binaries,
should be undertaken to further investigate these issues. 

\subsection{PSR J1806+2819}
\label{1806}
Initial timing observations of PSR J1806+2819 indicated that the barycentric spin period was varying between epochs,
presumably due to the Doppler effect as a result of its motion in a binary orbit with a
companion star. Initial measurements of the binary system's projected semi-major axis, orbital period, and time of
 periastron passage were determined by using \texttt{fit{\_}circular{\_}orbit.py} from the PRESTO software package, which models the varying spin period as an edge-on, circular orbit. These parameters, along with the orbital eccentricity and longitude of periastron, were then incorporated into the timing model. The results of the fits of these five Keplerian parameters, derived using the ELL1 binary model \citep{lcw01}, are shown in Table 3.

\begin{table}
\label{tb:bin}
\begin{center}
\caption{Timing-Derived Binary Parameters for PSR J1806$+$2819}
\begin{tabular}{ll}
\hline
Measured Parameters\\
\hline
Projected semi-major axis, $A_{1}$$\sin{i}$ (lt-s) & 21.608784(8) \\
Orbital eccentricity times, $\sin{\omega}$ $EPS1$  & $-$0.0000852(5)  \\
Orbital eccentricity times, $\cos{\omega}$ $EPS2$  & $-$0.0000192(7) \\
Orbital period, $P_{b}$ (days) & 43.866963(2) \\
Epoch of ascending node passage, $T_{asc}$ (MJD) & 57040.76929(2) \\
\hline
Derived Parameters\\
\hline
Orbital eccentricity, $e$ & 0.0000874(5) \\
Longitude of periastron argument, $\omega$ ($\deg$)& 257.3(4) \\
Time of periastron argument, $T_{0}$ (MJD) &  57072.13(5) \\
Mass function, $f_{M}$ ($\mathrm{M}_\odot$) & 0.005629873(6) \\
Minimum companion mass, $M_{c,min}$ ($\mathrm{M}_\odot$) & 0.25 \\ 

\hline \end{tabular}\end{center}
{\small {\bf Notes.} Projected pulsar semi-major axis of the orbit, orbital eccentricity multiplied by sine and cosine of the longitude of periastron argument, orbital period, and longitude of ascending node passage of PSR J1806$+$2819 measured using the ELL1 binary model. The orbital eccentricity, longitude of periastron argument, and time of periastron argument were derived from the measured quantities. The numbers in parentheses are the 1-$\sigma$ errors in the last digit reported by TEMPO.}
\end{table}

This pulsar is in a highly circular orbit of 44 days with a projected semi-major axis of 21.6 lt-s. Assuming a pulsar mass of 1.4 $\mathrm{M}_\odot$, we find that the minimum (where inclination, $i$ = 90\degr) and median (where inclination, $i$ = $60\degr$) companion masses are 0.25 $\mathrm{M}_\odot$ and 0.29 $\mathrm{M}_\odot$, respectively. The binary period companion mass relation \citep{ts99,imt16} gives a companion mass of 0.30 $\mathrm{M}_\odot$. This suggests a WD companion.

We searched for an optical counterpart to PSR J1806+2819 using data release 1 from the Panoramic Survey Telescope and Rapid Response System (PanSTARRS) 3$\pi$ survey \citep{cmm16}.  No source was present in the catalog at the position of the pulsar.  We searched the stacked images for each of the five bands (\texttt{grizy}) manually, and detect no source at the pulsar's position (see Figure 3). We use the average 5-$\sigma$ magnitude lower limits for the stacked survey data from \cite{cmm16} for this source: 23.3, 23.2, 23.1, 22.3, 21.4 for \texttt{grizy}, respectively. The estimated reddening of the source was determined using 3-D map of interstellar dust reddening by \cite{gsf15} at the DM-derived distance of 1.3 kpc. This was converted to an extinction in all five bands using Table 6 from \cite{sf11}.

\begin{figure}
         \begin{center}
	       	\includegraphics[width=0.5\textwidth]{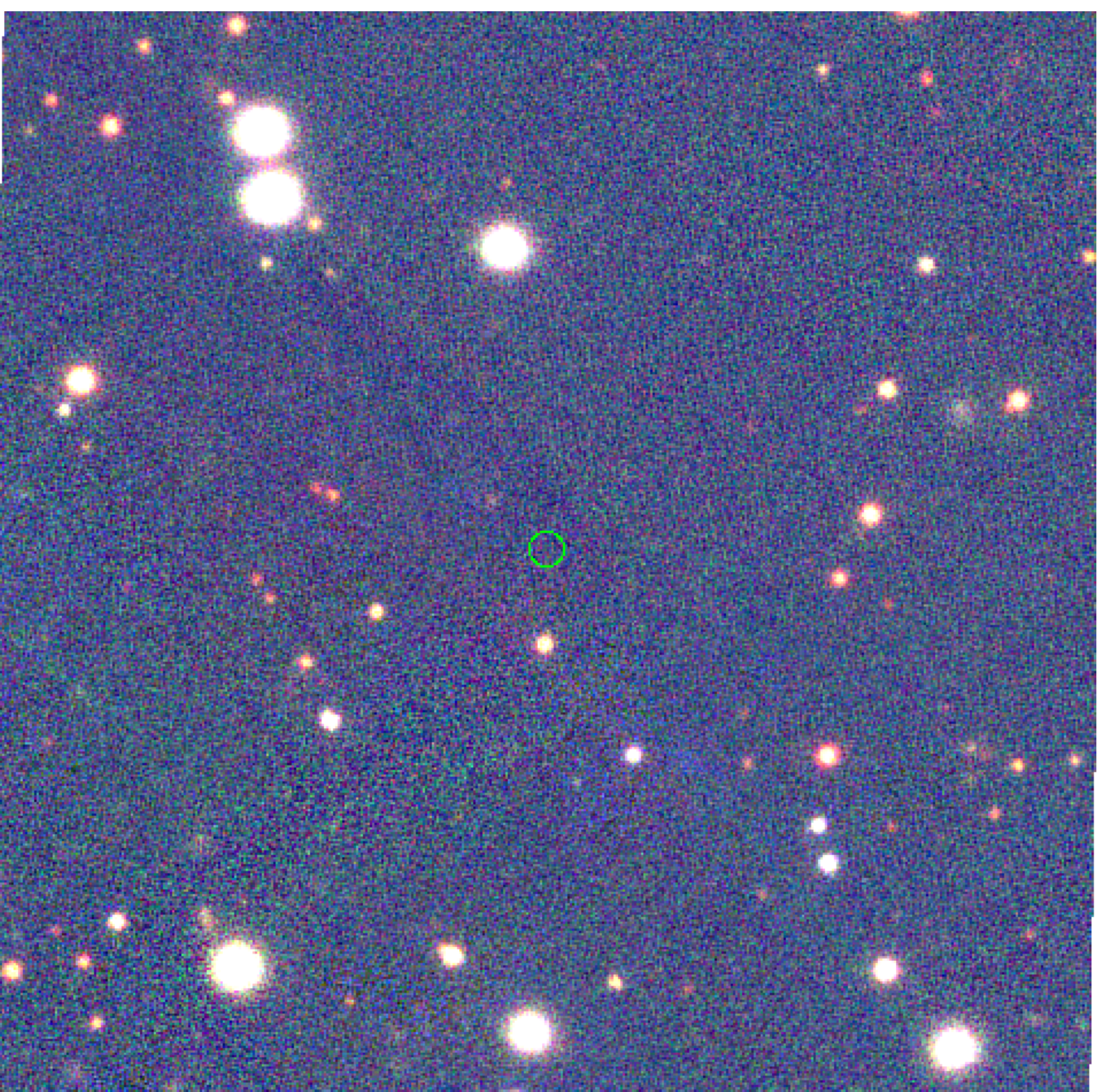}
       		\label{fig:gri}
        \end{center}
\caption{Three-color {\it gri} composite image from the PanSTARRS image cutout server at the position of PSR J1806+2819. This image is 2$\arcmin$ $\times$ 2$\arcmin$ in size where east is to the left and north is up. A 2$\arcsec$ error at the position of the pulsar is indicated by the circle. No optical counterpart was found for this pulsar. }
\end{figure}

We translated these limits into limits on the effective temperature and age of the presumed WD companion for an assumed companion mass of 0.30 $\mathrm{M}_\odot$. We first determined colors as a function of effective temperature for a 0.30 $\mathrm{M}_\odot$ WD using the Bergeron model\footnote[6]{http://www.astro.umontreal.ca/$\sim$bergeron/CoolingModels} \citep{hb06,ks06,tbg11,bwd11}. Note that the colors do not change significantly as a function of mass.  We used absolute normalizations from the models of \cite{imt16} appropriate for this mass value.  We find an upper limit to the effective temperature of 7100 K, with the most constraining limits being determined from the $r$ band.  We then used these limits to determine a lower limit on the cooling age of 1200 Myr.  For the minimum companion mass of 0.25 $\mathrm{M}_\odot$ (at an $i$ = $90\degr$), the effective temperature and cooling age limits are essentially unchanged.
 
PSR J1806+2819 is the only MSP presented in this paper. For MSPs to be considered for use in PTAs, the RMS timing residuals must be less than $\sim$1 $\mu$s. PSR J1806+2819 has a RMS value of 16.2 $\mu$s, which is too high to be considered for use in PTAs. There are numerous factors that could contribute to the high RMS value of PSR J1806+2819. The spin period for this pulsar (15 \rm{ms}) is larger than most MSPs in PTAs. The spin periods of MSPs used in the North American Nanohertz Observatory for Gravitational Waves (NANOGrav) nine-year data set range from 1.65 \rm{ms} to 16.05 \rm{ms} \citep{abb16}. If this pulsar was included in this PTA, it would have the second longest spin period. The large spin period of PSR J1806+2819 is not enough on its own to exclude it from PTA use; however, this pulsar also has a relatively wide pulse profile, shown in Figure \ref{fig:profiles}. The full width at half maximum (FWHM), listed in Table \ref{tb:prd}, is roughly 2 ms. This is much larger than the FWHM values for any of the MSPs in the NANOGrav PTA. 

One caveat to this discussion is that our timing measurements for PSR J1806$+$2819 were performed at 350 MHz, whereas most PTA timing is done around 1 GHz. The profile of PSR J1806$+$2819 could be narrower at 1 GHz and the RMS residual could decrease, but it is unlikely that these improvements would be sufficient to make it a suitable PTA addition. However, this pulsar is in the declination range visible to the Arecibo Observatory and should be tested with the very high sensitivity of Arecibo to see if it is suitable for inclusion in NANOGrav.

\subsection{PSR J0930$-$2301}
\label{0930}
The discovery observation of PSR J0930$-$2301 revealed that the pulse intensity was modulated, presumably due to pulse nulling. Pulse nulling is a phenomenon where the pulsed emission suddenly appears to drop to zero and then returns to its normal state \citep{b70}. Possible explanations for pulse nulling include: the pulsar undergoing complete cessation of emission \citep{klo06,gjk12}, the pulsar transitioning to weaker emission modes \citep{elg05,yws14}, the acceleration zone of the pulsar not being completely filled with electron-positron pairs \citep{dr01,jvl04}, and the pulsar beam of emission moving out of the line of sight from Earth \citep{dzg05,zgd07}. Mode changing is a related effect where the average pulse profile suddenly changes between two or more stable states.

Pulse nulling has been observed in about 100 pulsars \citep{gjk12}. The nulling fraction (NF), or fraction of the pulses with no detectable emission, can range from less than a percent to nearly 100\% \citep{r76,wmj07,gjw14}. The NF does not necessarily describe the duration of an individual null or the time between nulls. The null length is the timescale that the pulsar spends in the null state. This can range from a few pulse periods to years \citep{wmj07,llm12}. 

This pulsar was not detectable at many observed epochs possibly due to its nulling behavior, relative faintness, and spin frequency similar to harmonics of common sources of radio frequency interference (RFI). The high DM of PSR J0930$-$2301 makes it unlikely that scintillation played a role in these non-detections. After RFI excision was performed, using \texttt{rfifind} from the PRESTO software package, this pulsar was detected at enough epochs to achieve a full phase-connected timing solution.

To study the nulling behavior of PSR J0930$-$2301, a single pulse analysis was performed on a 56 minute timing observation. In offline processing, the dedispersed data were folded to 512 phase bins across the spin period to form subintegrations. Emission detected from PSR J0930$-$2301 was weak, so a subintegration length of 12 pulse periods was used to average enough pulses together to ensure a sufficient signal-to-noise ratio (\begin{math} \mathrm{S/N} \geq  5\end{math}). This analysis is insensitive to nulls less than or similar to the subintegration length. Therefore, the NF will be an underestimate and only the lower limit on this value can be determined.

We estimated the baseline using the off-pulse bins, and subtracted it from each subintegration to estimate the NF. Pulse energies of each subintegration were determined after subtraction for both the on-pulse and off-pulse regions. The pulse energies were then normalized by the mean energy of the on-pulse region. The summed energy of the on- and off-pulse regions was calculated using the same number of phase bins.

Figure 4 shows the histograms of the on- and off-pulse energy distributions, where the total number of bins for each energy distribution is equal to the number of subintegrations $N$. The excess at zero energy for the on-pulse energy distribution gives the fraction of nulled pulses for the pulsar \citep{r76}. The NF was calculated by first scaling the off-pulse energy distribution so that the number of pulses with energies less than zero was equal to that in the on-pulse energy distribution. Then we subtracted the on-pulse distribution from this scaled off-pulse distribution.  The NF is simply this scale factor.

\begin{figure}
         \begin{center}
                \includegraphics[width=0.5\textwidth]{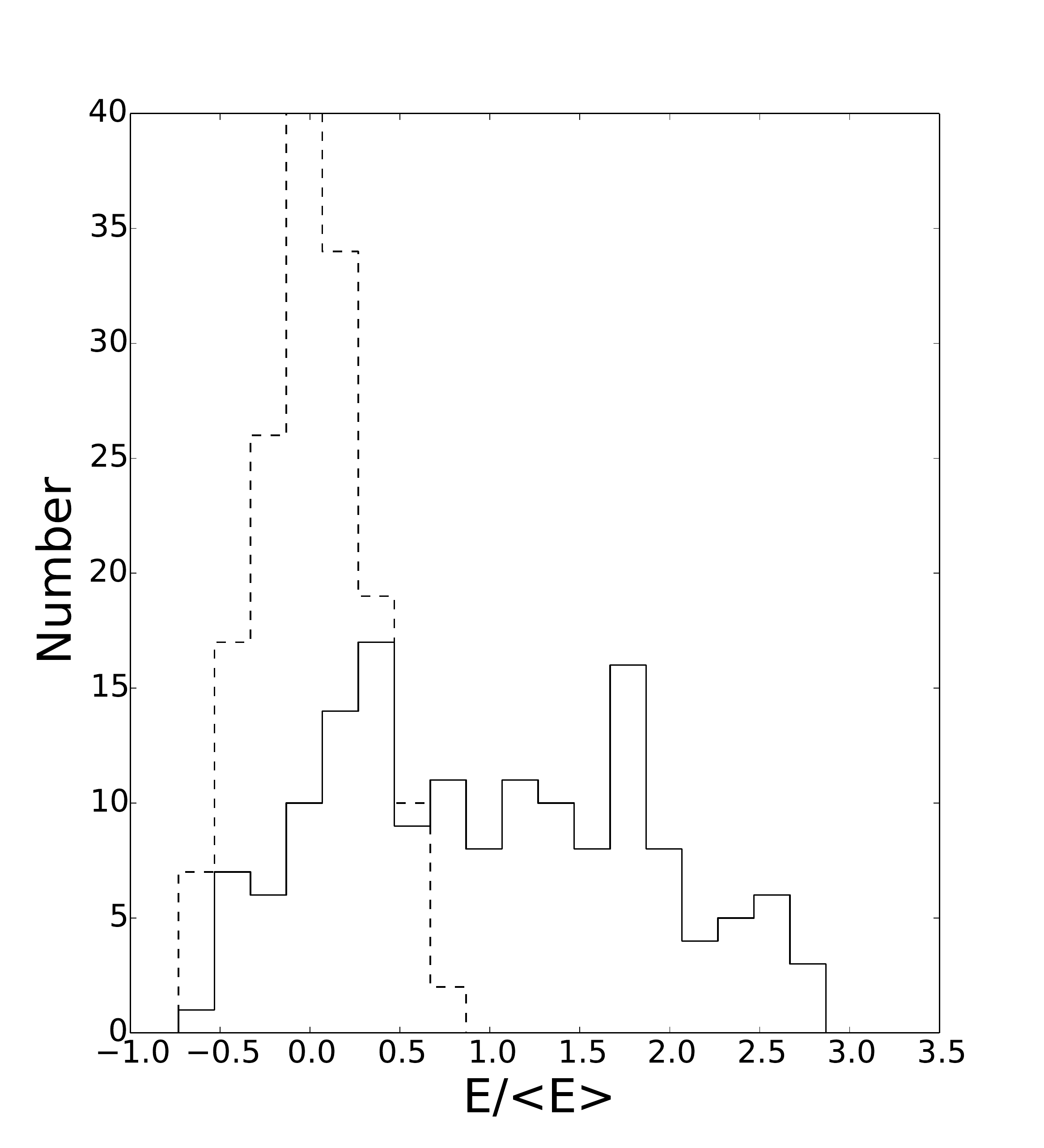}
        \label{fig:hist}
        \end{center}
\caption{Histogram of on-pulse (solid line) and off-pulse (dashed line) energies, after baseline subtraction, normalized by the mean pulse energy for PSR J0930$-$2301. The excess of the on-pulse energy distribution around zero energy gives the fraction of nulled pulses. }
\end{figure}

The null lengths were defined as the number of subsequent pulses all with pulse energies below a conservative threshold of five times the off-pulse RMS variation. Since the used subintegration length was 12 pulse periods, our analysis was insensitive to null lengths shorter than this duration. The overlap in the on-pulse and off-pulse histograms (Fig. 4) and the necessary integration of multiple pulses to obtain the required S/N mean we operate in different regime from high-precision nulling studies in which individual pulses are unambiguously identified (e.g. Figure 1 from \citealt{lkr02}). But these do allow us to place lower limits on the NF, and a qualitative estimate of the null lengths.

The lower limit of the NF for PSR J0930$-$2301 was estimated to be 30\%. The null lengths were found to be  36 pulse periods on average and ranged from 12 to 156 pulse periods. Given that the spin period of this pulsar is 1.8 \rm{s}, it can be concluded that it usually nulls for seconds up to minutes, with the longest null observed being roughly five minutes. This conclusion was supported via visual inspection of the phase$-$time diagram (Figure 5). Also, this figure shows that  the pulse profile has multiple components that turn on and off independently of each other. We conclude that this is evidence of mode changing.

\begin{figure}
         \begin{center}
                \includegraphics[width=0.5\textwidth]{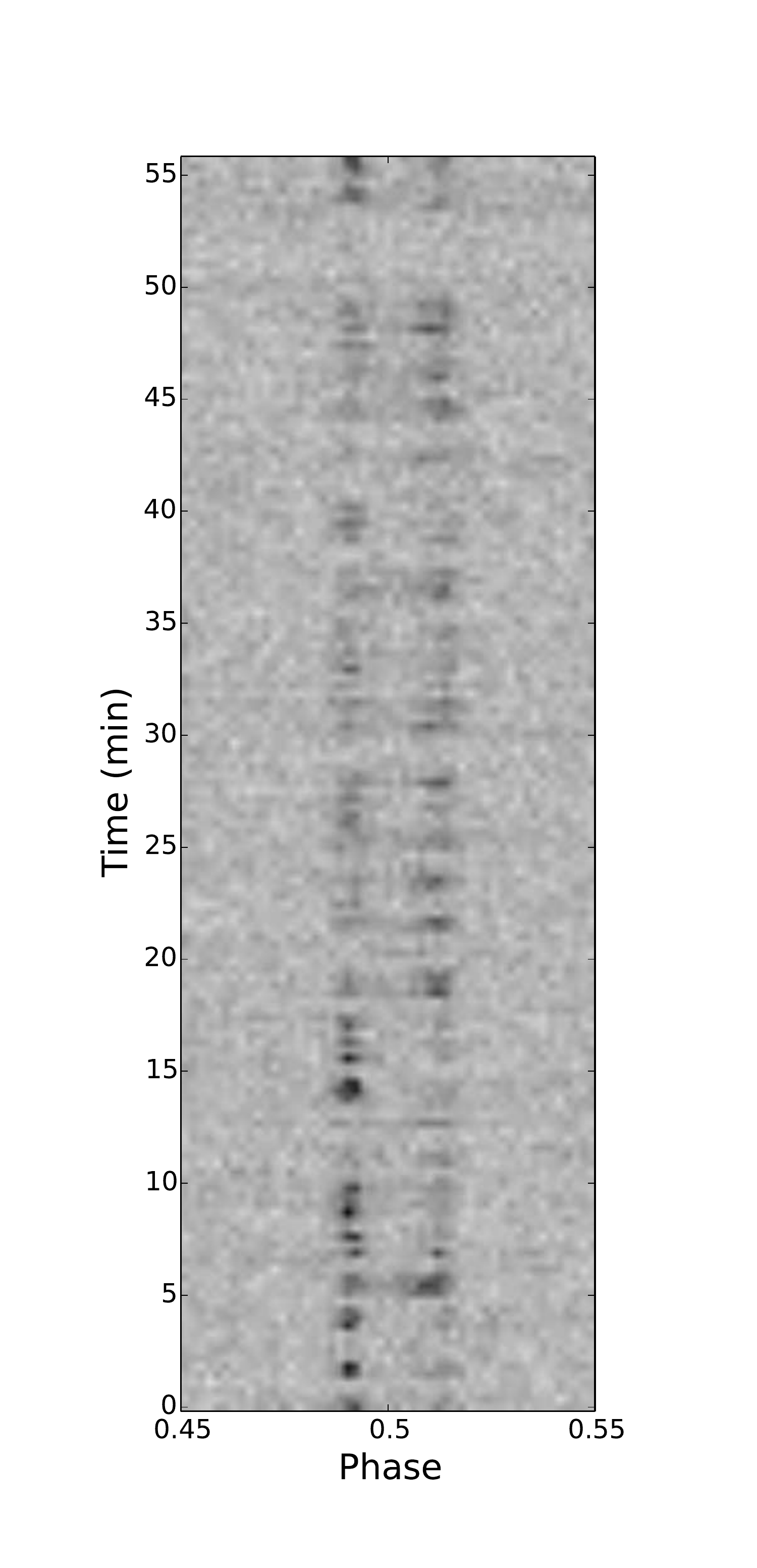}
        \label{fig:phasetime}
        \end{center}
\caption{Phase$-$time plot for PSR J0930$-$2301. One-tenth of the pulse period is shown and the grey-scale is linear intensity from zero in white to maximum intensity in black. Each subintegration is the average from 12 pulses. Both nulling and mode changing are seen.}
\end{figure}

\subsection{Electron Density Models}
\label{models}
As seen in the last two columns of Table 1, the distances estimated to the pulsars presented in this paper by the two most recent electron density models are very different for some pulsars. Since the YMW16 model is relatively new, these pulsars provide important data points on the relationship between the two models.

The NE2001 electron density model \citep{cl02} has been the standard tool to estimate distances to pulsars since 2001. This model incorporates 112 independent pulsar distances and 269 scattering measures to define an electron density model for the Galaxy. The components of the model include: thin and thick axisymmetric disks, spiral arms, a local arm, a local hot bubble around the Sun, super-bubbles in the first and third Galactic quadrants, and over-dense regions surrounding the Gum Nebula, the Vela supernova remnant, Galactic Loop I, and a small region around the Galactic Center. This model also incorporates clumps and voids in directions of known pulsars with DMs higher and lower, respectively, than predicted by the quasi-smooth component of the model.

The YMW16 \citep{ymw17} model has the same basic structure as the NE2001 model with some important differences. The YMW16 model utilizes 189 independent pulsar distances to define its electron density model. It incorporates many of the same Galactic components as the NE2001 model; however, it does not make use of interstellar scattering in building the model. Also, YMW16 does not attempt to correct individual pulsars with discrepant distances by adding clumps or voids in their respective directions. This type of feature is only added when a number of pulsars in a region have discrepant distances and/or there is independent evidence for this feature.

A majority of the distances estimated to the pulsars presented in this paper derived from the two models are significantly different. Neither model gives a reliable distance to PSR J0930$-$2301. The maximum DM predicted by the YMW16 model in the direction of PSR J0930$-$2301 is $\sim$60.8 pc cm$^{-3}$, which is lower than the measured DM of the pulsar (78.3 pc cm$^{-3}$). It is likely that this electron density model will significantly over-estimate the distances to pulsars in the direction of PSR J0930$-$2301. The maximum DM predicted by the  NE2001 model in the direction of PSR J0930$-$2301 is 77.12 pc cm$^{-3}$. This is closer to the timing-derived DM of the pulsar.

 The two models are in agreement for only one pulsar, assuming 25\% errors, PSR J1806+2819. This is the pulsar with the lowest DM of the ten presented in this paper, suggesting that distance discrepancies become greater with increasing DM. None of the pulsars presented in this paper are at high latitude (\begin{math} |b| > 40\degr \end{math}), where the NE2001 model has been found to systematically under-estimate distances \citep{lfl+06}. The distances derived using the YMW16 model are larger than the distances from the NE2001 for seven of the pulsars presented in this paper. The NE2001 model estimates a larger distance for only the closest pulsar, PSR J1119$-$2444.

\section{Conclusions and Future Work}
\label{future}
In this paper, we reported on the discovery and timing solutions of pulsars discovered through the Green Bank Northern Celestial Cap survey and the Pulsar Search Collaboratory. For each pulsar, a full phase-connected timing solution was achieved by measuring rotational and astrometric parameters.
PSR J0557$-$2948 is a partially recycled pulsar with no evidence of a binary companion. We conclude that this pulsar is likely the end result of a disrupted double neutron star system. PSR J1806$+$2819 is a MSP in a 44 day orbit with a WD companion. No optical counterpart was found, so lower limits were placed on the magnitude and age for the WD companion. This pulsar will not likely be considered for PTA use. We placed lower limits on the nulling fraction of PSR J0930$-$2301, a nulling pulsar, and also saw evidence of mode changing.
 
The GBNCC survey is currently 75\% complete, with an additional $\sim$ 50 pulsars expected to be discovered. Future observations of the pulsars discussed in Section \ref{conclusions}, could lead to additional results. Further timing observations and analysis of PSR J0557$-$2948 could result in a measurement of its proper motion. This would provide insight to the runaway velocity which neutron stars receive and the kick involved in the supernova explosions of their companions, and ultimately help better our understanding about how DNS systems and DRPs form. Higher frequency observations of PSR J1806$+$2819 could support including this pulsar in a PTA, depending on how significantly the RMS timing residuals is reduced. Future discoveries of pulsars with timing-derived DMs, and in particular, independent distances estimates will help to build electron density models of the Galaxy and lead to more accurate distance estimates to pulsars. 

\section*{Acknowledgements}
\label{ack}
The Green Bank Observatory is a facility of the National Science Foundation operated under cooperative agreement by Associated Universities, Inc.

We thank West Virginia University for its financial support of GBT operations, which enabled some of the observations for this project. 

MAM and AMK were supported by NSF awards AST-1327526 and OIA-1458952.
 
We thank Compute Canada, the McGill Center for High Performance Computing,
and Calcul Quebec for provision and maintenance of the
Guillimin supercomputer and related resources.

VMK receives support from an NSERC Discovery Grant, a Gerhard Herzberg Award, an R. Howard Webster Foundation Fellowship from the Canadian Institute for Advanced Research, the Canada Research Chairs Program, and the Lorne Trottier Chair in Astrophysics and Cosmology.

Pulsar work at UBC is supported by an NSERC Discovery Grant and by the Canadian Institute for Advanced Research.

MAM, DRL, JKS, FAJ, MED, DLK, KS, ALO, TDM, RS, ALM, SMR, and XS were supported by NSF Physics Frontiers Center award PtHY-1430284

JWTH and VIK acknowledge support from the European Research Council under the European Union's Seventh Framework Programme (FP/2007-2013) / ERC Grant Agreement nr. 337062.
\bibliography{modrefs,psrrefs}

\begin{thebibliography}{56}
\providecommand{\natexlab}[1]{#1}
\providecommand{\url}[1]{\texttt{#1}}
\expandafter\ifx\csname urlstyle\endcsname\relax
  \providecommand{\doi}[1]{doi: #1}\else
  \providecommand{\doi}{doi: \begingroup \urlstyle{rm}\Url}\fi

\bibitem[Alpar et~al.(1982)Alpar, Cheng, Ruderman, \& Shaham]{acrs82}
Alpar, M.~A., Cheng, A.~F., Ruderman, M.~A., \& Shaham, J., \emph{\nat}, 300,
  \penalty0 728--730, 1982.

\bibitem[{Archibald} et~al.(2009){Archibald}, {Stairs}, {Ransom}, {Kaspi},
  {Kondratiev}, {Lorimer}, {McLaughlin}, {Boyles}, {Hessels}, {Lynch}, {van
  Leeuwen}, {Roberts}, {Jenet}, {Champion}, {Rosen}, {Barlow}, {Dunlap}, \&
  {Remillard}]{asr09}
{Archibald}, A.~M., {Stairs}, I.~H., {Ransom}, S.~M., {Kaspi}, V.~M.,
  {Kondratiev}, V.~I., {Lorimer}, D.~R., {McLaughlin}, M.~A., {Boyles}, J.,
  {Hessels}, J.~W.~T., {Lynch}, R., {van Leeuwen}, J., {Roberts}, M.~S.~E.,
  {Jenet}, F., {Champion}, D.~J., {Rosen}, R., {Barlow}, B.~N., {Dunlap},
  B.~H., \& {Remillard}, R.~A., \emph{Science}, 324, \penalty0 1411, 2009.

\bibitem[{Arzoumanian} et~al.(2016){Arzoumanian}, {Brazier}, \&
  {Burke-Spolaor}]{abb16}
{Arzoumanian}, Z., {Brazier}, A., \& {Burke-Spolaor}, S., \emph{\apj}, 821,
  \penalty0 13, 2016.

\bibitem[{Arzoumanian} et~al.(2018){Arzoumanian}, {Baker}, {Brazier},
  {Burke-Spolaor}, {Chamberlin}, {Chatterjee}, {Christy}, {Cordes}, {Cornish},
  {Crawford}, {Thankful Cromartie}, {Crowter}, {DeCesar}, {Demorest}, {Dolch},
  {Ellis}, {Ferdman}, {Ferrara}, {Folkner}, {Fonseca}, {Garver-Daniels},
  {Gentile}, {Haas}, {Hazboun}, {Huerta}, {Islo}, {Jenet}, {Jones}, {Jones},
  {Kaplan}, {Kaspi}, {Lam}, {Lazio}, {Levin}, {Lommen}, {Lorimer}, {Luo},
  {Lynch}, {Madison}, {McLaughlin}, {McWilliams}, {Mingarelli}, {Ng}, {Nice},
  {Park}, {Pennucci}, {Pol}, {Ransom}, {Ray}, {Rasskazov}, {Siemens}, {Simon},
  {Spiewak}, {Stairs}, {Stinebring}, {Stovall}, {Swiggum}, {Taylor},
  {Vallisneri}, {Vigeland}, \& {Zhu}]{abb18}
{Arzoumanian}, Z., {Baker}, P.~T., {Brazier}, A., {Burke-Spolaor}, S.,
  {Chamberlin}, S.~J., {Chatterjee}, S., {Christy}, B., {Cordes}, J.~M.,
  {Cornish}, N.~J., {Crawford}, F., {Thankful Cromartie}, H., {Crowter}, K.,
  {DeCesar}, M., {Demorest}, P.~B., {Dolch}, T., {Ellis}, J.~A., {Ferdman},
  R.~D., {Ferrara}, E., {Folkner}, W.~M., {Fonseca}, E., {Garver-Daniels}, N.,
  {Gentile}, P.~A., {Haas}, R., {Hazboun}, J.~S., {Huerta}, E.~A., {Islo}, K.,
  {Jenet}, F., {Jones}, G., {Jones}, M.~L., {Kaplan}, D.~L., {Kaspi}, V.~M.,
  {Lam}, M.~T., {Lazio}, T.~J.~W., {Levin}, L., {Lommen}, A.~N., {Lorimer},
  D.~R., {Luo}, J., {Lynch}, R.~S., {Madison}, D.~R., {McLaughlin}, M.~A.,
  {McWilliams}, S.~T., {Mingarelli}, C.~M.~F., {Ng}, C., {Nice}, D.~J., {Park},
  R.~S., {Pennucci}, T.~T., {Pol}, N.~S., {Ransom}, S.~M., {Ray}, P.~S.,
  {Rasskazov}, A., {Siemens}, X., {Simon}, J., {Spiewak}, R., {Stairs}, I.~H.,
  {Stinebring}, D.~R., {Stovall}, K., {Swiggum}, J., {Taylor}, S.~R.,
  {Vallisneri}, M., {Vigeland}, S., \& {Zhu}, W.~W., \emph{ArXiv e-prints},
  2018.

\bibitem[{Backer}(1970)]{b70}
{Backer}, D.~C., \emph{\nat}, 228, \penalty0 42--43, 1970.

\bibitem[Bergeron et~al.(2011)Bergeron, Wesemael, Dufour, Beauchamp, Hunter,
  Saffer, Gianninas, Ruiz, Limoges, Dufour, Fontaine, \& Liebert]{bwd11}
Bergeron, P., Wesemael, F., Dufour, P., Beauchamp, A., Hunter, C., Saffer,
  R.~A., Gianninas, A., Ruiz, M.~T., Limoges, M.-M., Dufour, P., Fontaine, G.,
  \& Liebert, J., \emph{\aj}, 737\penalty0 (1), \penalty0 28, 2011.

\bibitem[{Bhattacharya} \& {van den Heuvel}(1991)]{bh91}
{Bhattacharya}, D. \& {van den Heuvel}, E.~P.~J., \emph{\physrep}, 203,
  \penalty0 1--124, 1991.

\bibitem[Camilo et~al.(1993)Camilo, Nice, \& Taylor]{cnt93}
Camilo, F., Nice, D.~J., \& Taylor, J.~H., \emph{\apjl}, 412, \penalty0
  L37--L40, 1993.

\bibitem[{Chambers} et~al.(2016){Chambers}, {Magnier}, {Metcalfe},
  {Flewelling}, {Huber}, {Waters}, {Denneau}, {Draper}, {Farrow}, {Finkbeiner},
  {Holmberg}, {Koppenhoefer}, {Price}, {Saglia}, {Schlafly}, {Smartt},
  {Sweeney}, {Wainscoat}, {Burgett}, {Grav}, {Heasley}, {Hodapp}, {Jedicke},
  {Kaiser}, {Kudritzki}, {Luppino}, {Lupton}, {Monet}, {Morgan}, {Onaka},
  {Stubbs}, {Tonry}, {Banados}, {Bell}, {Bender}, {Bernard}, {Botticella},
  {Casertano}, {Chastel}, {Chen}, {Chen}, {Cole}, {Deacon}, {Frenk},
  {Fitzsimmons}, {Gezari}, {Goessl}, {Goggia}, {Goldman}, {Grebel}, {Hambly},
  {Hasinger}, {Heavens}, {Heckman}, {Henderson}, {Henning}, {Holman}, {Hopp},
  {Ip}, {Isani}, {Keyes}, {Koekemoer}, {Kotak}, {Long}, {Lucey}, {Liu},
  {Martin}, {McLean}, {Morganson}, {Murphy}, {Nieto-Santisteban}, {Norberg},
  {Peacock}, {Pier}, {Postman}, {Primak}, {Rae}, {Rest}, {Riess}, {Riffeser},
  {Rix}, {Roser}, {Schilbach}, {Schultz}, {Scolnic}, {Szalay}, {Seitz},
  {Shiao}, {Small}, {Smith}, {Soderblom}, {Taylor}, {Thakar}, {Thiel},
  {Thilker}, {Urata}, {Valenti}, {Walter}, {Watters}, {Werner}, {White},
  {Wood-Vasey}, \& {Wyse}]{cmm16}
{Chambers}, K.~C., {Magnier}, E.~A., {Metcalfe}, N., {Flewelling}, H.~A.,
  {Huber}, M.~E., {Waters}, C.~Z., {Denneau}, L., {Draper}, P.~W., {Farrow},
  D., {Finkbeiner}, D.~P., {Holmberg}, C., {Koppenhoefer}, J., {Price}, P.~A.,
  {Saglia}, R.~P., {Schlafly}, E.~F., {Smartt}, S.~J., {Sweeney}, W.,
  {Wainscoat}, R.~J., {Burgett}, W.~S., {Grav}, T., {Heasley}, J.~N., {Hodapp},
  K.~W., {Jedicke}, R., {Kaiser}, N., {Kudritzki}, R.-P., {Luppino}, G.~A.,
  {Lupton}, R.~H., {Monet}, D.~G., {Morgan}, J.~S., {Onaka}, P.~M., {Stubbs},
  C.~W., {Tonry}, J.~L., {Banados}, E., {Bell}, E.~F., {Bender}, R., {Bernard},
  E.~J., {Botticella}, M.~T., {Casertano}, S., {Chastel}, S., {Chen}, W.-P.,
  {Chen}, X., {Cole}, S., {Deacon}, N., {Frenk}, C., {Fitzsimmons}, A.,
  {Gezari}, S., {Goessl}, C., {Goggia}, T., {Goldman}, B., {Grebel}, E.~K.,
  {Hambly}, N.~C., {Hasinger}, G., {Heavens}, A.~F., {Heckman}, T.~M.,
  {Henderson}, R., {Henning}, T., {Holman}, M., {Hopp}, U., {Ip}, W.-H.,
  {Isani}, S., {Keyes}, C.~D., {Koekemoer}, A., {Kotak}, R., {Long}, K.~S.,
  {Lucey}, J.~R., {Liu}, M., {Martin}, N.~F., {McLean}, B., {Morganson}, E.,
  {Murphy}, D.~N.~A., {Nieto-Santisteban}, M.~A., {Norberg}, P., {Peacock},
  J.~A., {Pier}, E.~A., {Postman}, M., {Primak}, N., {Rae}, C., {Rest}, A.,
  {Riess}, A., {Riffeser}, A., {Rix}, H.~W., {Roser}, S., {Schilbach}, E.,
  {Schultz}, A.~S.~B., {Scolnic}, D., {Szalay}, A., {Seitz}, S., {Shiao}, B.,
  {Small}, E., {Smith}, K.~W., {Soderblom}, D., {Taylor}, A.~N., {Thakar},
  A.~R., {Thiel}, J., {Thilker}, D., {Urata}, Y., {Valenti}, J., {Walter}, F.,
  {Watters}, S.~P., {Werner}, S., {White}, R., {Wood-Vasey}, W.~M., \& {Wyse},
  R., \emph{ArXiv e-prints}, 2016.

\bibitem[{Cordes} \& {Lazio}(2002)]{cl02}
{Cordes}, J.~M. \& {Lazio}, T.~J.~W., \emph{ArXiv Astrophysics e-prints}, 2002.

\bibitem[Damour \& Taylor(1992)]{dt92}
Damour, T. \& Taylor, J.~H., 45, \penalty0 1840--1868, 1992.

\bibitem[{Deneva} et~al.(2016){Deneva}, {Stovall}, {McLaughlin}, {Bagchi},
  {Bates}, {Freire}, {Martinez}, {Jenet}, \& {Garver-Daniels}]{dsm16}
{Deneva}, J.~S., {Stovall}, K., {McLaughlin}, M.~A., {Bagchi}, M., {Bates},
  S.~D., {Freire}, P.~C.~C., {Martinez}, J.~G., {Jenet}, F., \&
  {Garver-Daniels}, N., \emph{\apj}, 821, \penalty0 10, 2016.

\bibitem[Deshpande \& Rankin(2001)]{dr01}
Deshpande, A.~A. \& Rankin, J.~M., \emph{\mnras}, 322, \penalty0 438--460,
  2001.

\bibitem[{Dyks} et~al.(2005){Dyks}, {Zhang}, \& {Gil}]{dzg05}
{Dyks}, J., {Zhang}, B., \& {Gil}, J., \emph{\apjl}, 626, \penalty0 L45--L47,
  2005.

\bibitem[{Esamdin} et~al.(2005){Esamdin}, {Lyne}, {Graham-Smith}, {Kramer},
  {Manchester}, \& {Wu}]{elg05}
{Esamdin}, A., {Lyne}, A.~G., {Graham-Smith}, F., {Kramer}, M., {Manchester},
  R.~N., \& {Wu}, X., \emph{\mnras}, 356, \penalty0 59--65, 2005.

\bibitem[{Folkner} et~al.(2014){Folkner}, {Williams}, {Boggs}, {Park}, \&
  {Kuchynka}]{fwb14}
{Folkner}, W.~M., {Williams}, J.~G., {Boggs}, D.~H., {Park}, R.~S., \&
  {Kuchynka}, P., \emph{Interplanetary Network Progress Report}, 196, \penalty0
  1--81, 2014.

\bibitem[Fruchter et~al.(1988)Fruchter, Stinebring, \& Taylor]{fst88}
Fruchter, A.~S., Stinebring, D.~R., \& Taylor, J.~H., \emph{\nat}, 333,
  \penalty0 237--239, 1988.

\bibitem[{Gajjar} et~al.(2012){Gajjar}, {Joshi}, \& {Kramer}]{gjk12}
{Gajjar}, V., {Joshi}, B.~C., \& {Kramer}, M., \emph{\mnras}, 424, \penalty0
  1197--1205, 2012.

\bibitem[{Gajjar} et~al.(2014){Gajjar}, {Joshi}, \& {Wright}]{gjw14}
{Gajjar}, V., {Joshi}, B.~C., \& {Wright}, G., \emph{\mnras}, 439, \penalty0
  221--233, 2014.

\bibitem[{Green} et~al.(2015){Green}, {Schlafly}, {Finkbeiner}, {Rix},
  {Martin}, {Burgett}, {Draper}, {Flewelling}, {Hodapp}, {Kaiser}, {Kudritzki},
  {Magnier}, {Metcalfe}, {Price}, {Tonry}, \& {Wainscoat}]{gsf15}
{Green}, G.~M., {Schlafly}, E.~F., {Finkbeiner}, D.~P., {Rix}, H.-W., {Martin},
  N., {Burgett}, W., {Draper}, P.~W., {Flewelling}, H., {Hodapp}, K., {Kaiser},
  N., {Kudritzki}, R.~P., {Magnier}, E., {Metcalfe}, N., {Price}, P., {Tonry},
  J., \& {Wainscoat}, R., \emph{\apj}, 810:\penalty0 25, 2015.

\bibitem[{Hills}(1983)]{h83}
{Hills}, J.~G., \emph{\apj}, 267, \penalty0 322--333, 1983.

\bibitem[{Holberg} \& {Bergeron}(2006)]{hb06}
{Holberg}, J.~B. \& {Bergeron}, P., \emph{\aj}, 132, \penalty0 1221--1233,
  2006.

\bibitem[{Istrate} et~al.(2016){Istrate}, {Marchant}, {Tauris}, {Langer},
  {Stancliffe}, \& {Grassitelli}]{imt16}
{Istrate}, A.~G., {Marchant}, P., {Tauris}, T.~M., {Langer}, N., {Stancliffe},
  R.~J., \& {Grassitelli}, L., \emph{\aap}, 595:\penalty0 A35, 2016.

\bibitem[{Janssen} \& {van Leeuwen}(2004)]{jvl04}
{Janssen}, G.~H. \& {van Leeuwen}, J., \emph{\aap}, 425, \penalty0 255--261,
  2004.

\bibitem[{Kowalski} \& {Saumon}(2006)]{ks06}
{Kowalski}, P.~M. \& {Saumon}, D., \emph{\apjl}, 651, \penalty0 L137--L140,
  2006.

\bibitem[{Kramer} et~al.(2006){Kramer}, {Lyne}, {O'Brien}, {Jordan}, \&
  {Lorimer}]{klo06}
{Kramer}, M., {Lyne}, A.~G., {O'Brien}, J.~T., {Jordan}, C.~A., \& {Lorimer},
  D.~R., \emph{Science}, 312, \penalty0 549--551, 2006.

\bibitem[Lange et~al.(2001)Lange, Camilo, Wex, Kramer, Backer, Lyne, \&
  Doroshenko]{lcw01}
Lange, C., Camilo, F., Wex, N., Kramer, M., Backer, D., Lyne, A., \&
  Doroshenko, O., \emph{\mnras}, 326\penalty0 (1), \penalty0 274, 2001.

\bibitem[{Lentati} et~al.(2015){Lentati}, {Taylor}, \& {Mingarelli}]{ltm15}
{Lentati}, L., {Taylor}, S.~R., \& {Mingarelli}, C.~M.~F., \emph{\mnras}, 453,
  \penalty0 2576--2598, 2015.

\bibitem[{Lorimer} et~al.(2004){Lorimer}, {McLaughlin}, {Arzoumanian},
  {Xilouris}, {Cordes}, {Lommen}, {Fruchter}, {Chandler}, \& {Backer}]{lma04}
{Lorimer}, D.~R., {McLaughlin}, M.~A., {Arzoumanian}, Z., {Xilouris}, K.~M.,
  {Cordes}, J.~M., {Lommen}, A.~N., {Fruchter}, A.~S., {Chandler}, A.~M., \&
  {Backer}, D.~C., \emph{\mnras}, 347, \penalty0 L21--L25, 2004.

\bibitem[{Lorimer} et~al.(2006){Lorimer}, {Faulkner}, {Lyne}, {Manchester},
  {Kramer}, {McLaughlin}, {Hobbs}, {Possenti}, {Stairs}, {Camilo}, {Burgay},
  {D'Amico}, {Corongiu}, \& {Crawford}]{lfl+06}
{Lorimer}, D.~R., {Faulkner}, A.~J., {Lyne}, A.~G., {Manchester}, R.~N.,
  {Kramer}, M., {McLaughlin}, M.~A., {Hobbs}, G., {Possenti}, A., {Stairs},
  I.~H., {Camilo}, F., {Burgay}, M., {D'Amico}, N., {Corongiu}, A., \&
  {Crawford}, F., \emph{\mnras}, 372, \penalty0 777--800, 2006.

\bibitem[{Lorimer} et~al.(2012){Lorimer}, {Lyne}, {McLaughlin}, {Kramer},
  {Pavlov}, \& {Chang}]{llm12}
{Lorimer}, D.~R., {Lyne}, A.~G., {McLaughlin}, M.~A., {Kramer}, M., {Pavlov},
  G.~G., \& {Chang}, C., \emph{\apj}, 758:\penalty0 141, 2012.

\bibitem[Lorimer(2008)]{l08}
Lorimer, D.~R., \emph{Living Reviews in Relativity}, 11\penalty0 (1), \penalty0
  8, 2008.

\bibitem[{Manchester} et~al.(2005){Manchester}, {Hobbs}, {Teoh}, \&
  {Hobbs}]{mhth05}
{Manchester}, R.~N., {Hobbs}, G.~B., {Teoh}, A., \& {Hobbs}, M., \emph{\aj},
  129, \penalty0 1993--2006, 2005.

\bibitem[{McLaughlin} et~al.(2006){McLaughlin}, {Lyne}, {Lorimer}, {Kramer},
  {Faulkner}, {Manchester}, {Cordes}, {Camilo}, {Possenti}, {Stairs}, {Hobbs},
  {D'Amico}, {Burgay}, \& {O'Brien}]{mll06}
{McLaughlin}, M.~A., {Lyne}, A.~G., {Lorimer}, D.~R., {Kramer}, M., {Faulkner},
  A.~J., {Manchester}, R.~N., {Cordes}, J.~M., {Camilo}, F., {Possenti}, A.,
  {Stairs}, I.~H., {Hobbs}, G., {D'Amico}, N., {Burgay}, M., \& {O'Brien},
  J.~T., \emph{\nat}, 439, \penalty0 817--820, 2006.

\bibitem[Nice \& Taylor(1995)]{nt95}
Nice, D.~J. \& Taylor, J.~H., 441, \penalty0 429--435, 1995.

\bibitem[{Portegies Zwart} et~al.(2011){Portegies Zwart}, {van den Heuvel},
  {van Leeuwen}, \& {Nelemans}]{phl11}
{Portegies Zwart}, S., {van den Heuvel}, E.~P.~J., {van Leeuwen}, J., \&
  {Nelemans}, G., \emph{\apj}, 734:\penalty0 55, 2011.

\bibitem[{Ransom}(2011)]{rs11}
{Ransom}, S.
\newblock \emph{{PRESTO: PulsaR Exploration and Search TOolkit}}, 2011.

\bibitem[{Ransom} et~al.(2009){Ransom}, {Demorest}, {Ford}, {McCullough},
  {Ray}, {DuPlain}, \& {Brandt}]{rdf09}
{Ransom}, S.~M., {Demorest}, P., {Ford}, J., {McCullough}, R., {Ray}, J.,
  {DuPlain}, R., \& {Brandt}, P.
\newblock \emph{{GUPPI: Green Bank Ultimate Pulsar Processing Instrument}}.
\newblock In \emph{American Astronomical Society Meeting Abstracts \#214},
  volume 214 of \emph{American Astronomical Society Meeting Abstracts}, page
  605.08, 2009.

\bibitem[{Ritchings}(1976)]{r76}
{Ritchings}, R.~T., \emph{\mnras}, 176, \penalty0 249--263, 1976.

\bibitem[{Roberts}(2011)]{r11}
{Roberts}, M.~S.~E.
\newblock \emph{{New Black Widows and Redbacks in the Galactic Field}}.
\newblock In {Burgay}, M., {D'Amico}, N., {Esposito}, P., {Pellizzoni}, A., \&
  {Possenti}, A., editors, \emph{American Institute of Physics Conference
  Series}, volume 1357 of \emph{American Institute of Physics Conference
  Series}, pages 127--130, 2011.

\bibitem[{Rosen} et~al.(2013){Rosen}, {Swiggum}, {McLaughlin}, {Lorimer},
  {Yun}, {Heatherly}, {Boyles}, {Lynch}, {Kondratiev}, {Scoles}, {Ransom},
  {Moniot}, {Cottrill}, {Weaver}, {Snider}, {Thompson}, {Raycraft},
  {Dudenhoefer}, {Allphin}, {Thorley}, {Meadows}, {Marchiny}, {Liska},
  {O'Dwyer}, {Butler}, {Bloxton}, {Mabry}, {Abate}, {Boothe}, {Pritt},
  {Alberth}, {Green}, {Crowley}, {Agee}, {Nagley}, {Sargent}, {Hinson},
  {Smith}, {McNeely}, {Quigley}, {Pennington}, {Chen}, {Maynard}, {Loope},
  {Bielski}, {McGough}, {Gural}, {Colvin}, {Tso}, {Ewen}, {Zhang},
  {Ciccarella}, {Bukowski}, {Novotny}, {Gore}, {Sarver}, {Johnson},
  {Cunningham}, {Collins}, {Gardner}, {Monteleone}, {Hall}, {Schweinhagen},
  {Ayers}, {Jay}, {Uosseph}, {Dunkum}, {Pal}, {Dydiw}, {Sterling}, \&
  {Phan}]{rsm13}
{Rosen}, R., {Swiggum}, J., {McLaughlin}, M.~A., {Lorimer}, D.~R., {Yun}, M.,
  {Heatherly}, S.~A., {Boyles}, J., {Lynch}, R., {Kondratiev}, V.~I., {Scoles},
  S., {Ransom}, S.~M., {Moniot}, M.~L., {Cottrill}, A., {Weaver}, M., {Snider},
  A., {Thompson}, C., {Raycraft}, M., {Dudenhoefer}, J., {Allphin}, L.,
  {Thorley}, J., {Meadows}, B., {Marchiny}, G., {Liska}, A., {O'Dwyer}, A.~M.,
  {Butler}, B., {Bloxton}, S., {Mabry}, H., {Abate}, H., {Boothe}, J., {Pritt},
  S., {Alberth}, J., {Green}, A., {Crowley}, R.~J., {Agee}, A., {Nagley}, S.,
  {Sargent}, N., {Hinson}, E., {Smith}, K., {McNeely}, R., {Quigley}, H.,
  {Pennington}, A., {Chen}, S., {Maynard}, T., {Loope}, L., {Bielski}, N.,
  {McGough}, J.~R., {Gural}, J.~C., {Colvin}, S., {Tso}, S., {Ewen}, Z.,
  {Zhang}, M., {Ciccarella}, N., {Bukowski}, B., {Novotny}, C.~B., {Gore}, J.,
  {Sarver}, K., {Johnson}, S., {Cunningham}, H., {Collins}, D., {Gardner}, D.,
  {Monteleone}, A., {Hall}, J., {Schweinhagen}, R., {Ayers}, J., {Jay}, S.,
  {Uosseph}, B., {Dunkum}, D., {Pal}, J., {Dydiw}, S., {Sterling}, M., \&
  {Phan}, E., \emph{\apj}, 768:\penalty0 85, 2013.

\bibitem[{Schlafly} \& {Finkbeiner}(2011)]{sf11}
{Schlafly}, E.~F. \& {Finkbeiner}, D.~P., \emph{\apj}, 737:\penalty0 103, 2011.

\bibitem[Shannon et~al.(2013)Shannon, Ravi, Coles, Hobbs, Keith, Manchester,
  Wyithe, Bailes, Bhat, Burke-Spolaor, Khoo, Levin, Os{\l}owski, Sarkissian,
  van Straten, Verbiest, \& Wang]{srm13}
Shannon, R.~M., Ravi, V., Coles, W.~A., Hobbs, G., Keith, M.~J., Manchester,
  R.~N., Wyithe, J. S.~B., Bailes, M., Bhat, N. D.~R., Burke-Spolaor, S., Khoo,
  J., Levin, Y., Os{\l}owski, S., Sarkissian, J.~M., van Straten, W., Verbiest,
  J. P.~W., \& Wang, J.-B., \emph{Science}, 342\penalty0 (6156), \penalty0
  334--337, 2013.

\bibitem[{Shklovskii}(1970)]{s70}
{Shklovskii}, I.~S., \emph{\sovast}, 13, \penalty0 562, 1970.

\bibitem[{Stovall} et~al.(2014){Stovall}, {Lynch}, \& {Ransom}]{slr14}
{Stovall}, K., {Lynch}, R.~S., \& {Ransom}, S.~M., \emph{\apj}, 791, \penalty0
  67, 2014.

\bibitem[{Swiggum} et~al.(2015){Swiggum}, {Rosen}, {McLaughlin}, {Lorimer},
  {Heatherly}, {Lynch}, {Scoles}, {Hockett}, {Filik}, {Marlowe}, {Barlow},
  {Weaver}, {Hilzendeger}, {Ernst}, {Crowley}, {Stone}, {Miller}, {Nunez},
  {Trevino}, {Doehler}, {Cramer}, {Yencsik}, {Thorley}, {Andrews}, {Laws},
  {Wenger}, {Teter}, {Snyder}, {Dittmann}, {Gray}, {Carter}, {McGough},
  {Dydiw}, {Pruett}, {Fink}, \& {Vanderhout}]{srm15}
{Swiggum}, J.~K., {Rosen}, R., {McLaughlin}, M.~A., {Lorimer}, D.~R.,
  {Heatherly}, S., {Lynch}, R., {Scoles}, S., {Hockett}, T., {Filik}, E.,
  {Marlowe}, J.~A., {Barlow}, B.~N., {Weaver}, M., {Hilzendeger}, M., {Ernst},
  S., {Crowley}, R., {Stone}, E., {Miller}, B., {Nunez}, R., {Trevino}, G.,
  {Doehler}, M., {Cramer}, A., {Yencsik}, D., {Thorley}, J., {Andrews}, R.,
  {Laws}, A., {Wenger}, K., {Teter}, L., {Snyder}, T., {Dittmann}, A., {Gray},
  S., {Carter}, M., {McGough}, C., {Dydiw}, S., {Pruett}, C., {Fink}, J., \&
  {Vanderhout}, A., \emph{\apj}, 805:\penalty0 156, 2015.

\bibitem[{Tauris} \& {Savonije}(1999)]{ts99}
{Tauris}, T.~M. \& {Savonije}, G.~J., \emph{\aap}, 350, \penalty0 928--944,
  1999.

\bibitem[{Tauris} \& {Takens}(1998)]{tt98}
{Tauris}, T.~M. \& {Takens}, R.~J., \emph{\aap}, 330, \penalty0 1047--1059,
  1998.

\bibitem[{Tauris} et~al.(2012){Tauris}, {Langer}, \& {Kramer}]{tlm12}
{Tauris}, T.~M., {Langer}, N., \& {Kramer}, M., \emph{\mnras}, 425, \penalty0
  1601--1627, 2012.

\bibitem[{Taylor}(1992)]{t93}
{Taylor}, J.~H., \emph{Philosophical Transactions of the Royal Society of
  London Series A}, 341, \penalty0 117--134, 1992.

\bibitem[{Tremblay} et~al.(2011){Tremblay}, {Bergeron}, \& {Gianninas}]{tbg11}
{Tremblay}, P.-E., {Bergeron}, P., \& {Gianninas}, A., \emph{\apj},
  730:\penalty0 128, 2011.

\bibitem[{van Leeuwen} et~al.(2002){van Leeuwen}, {Kouwenhoven},
  {Ramachandran}, {Rankin}, \& {Stappers}]{lkr02}
{van Leeuwen}, A.~G.~J., {Kouwenhoven}, M.~L.~A., {Ramachandran}, R., {Rankin},
  J.~M., \& {Stappers}, B.~W., \emph{\aap}, 387, \penalty0 169--178, 2002.

\bibitem[Wang et~al.(2007)Wang, Manchester, \& Johnston]{wmj07}
Wang, N., Manchester, R.~N., \& Johnston, S., \emph{\mnras}, 377\penalty0 (3),
  \penalty0 1383, 2007.

\bibitem[{Yao} et~al.(2017){Yao}, {Manchester}, \& {Wang}]{ymw17}
{Yao}, J.~M., {Manchester}, R.~N., \& {Wang}, N., \emph{\apj}, 835, \penalty0
  29, 2017.

\bibitem[{Young} et~al.(2014){Young}, {Weltevrede}, {Stappers}, {Lyne}, \&
  {Kramer}]{yws14}
{Young}, N.~J., {Weltevrede}, P., {Stappers}, B.~W., {Lyne}, A.~G., \&
  {Kramer}, M., \emph{\mnras}, 442, \penalty0 2519--2533, 2014.

\bibitem[{Zhang} et~al.(2007){Zhang}, {Gil}, \& {Dyks}]{zgd07}
{Zhang}, B., {Gil}, J., \& {Dyks}, J., \emph{\mnras}, 374, \penalty0
  1103--1107, 2007.

\end{thebibliography}
\bibliographystyle{bibstyle}

\end{document}